\documentclass[preprint]{elsarticle}

\usepackage{amssymb}
\usepackage{amsmath}
\usepackage{subfigure}
\usepackage{hyperref}
\journal{ }

\begin{document}
\begin{frontmatter}

\title{Study Majorana Neutrino Contribution to B-meson Semi-leptonic Rare Decays}
\author[sdu]{Ying Wang}
\ead{wang\_y@mail.sdu.edu.cn}
\author[sdu]{Shou-Shan Bao}
\ead{ssbao@sdu.edu.cn}
\author[ytu]{Zuo-Hong Li}
\ead{lizh@ytu.edu.cn}
\author[ytu]{Nan Zhu}
\ead{zhunan426@126.com}
\author[sdu]{Zong-Guo Si}
\ead{zgsi@sdu.edu.cn}

\address[sdu]{School of Physics, Shandong University, Jinan 250100, P. R. China}
\address[ytu]{Department of Physics, Yantai University, Yantai 264005, P. R. China}

\begin{abstract}
B meson semi-leptonic rare decays are sensitive to new physics beyond standard model.
We study the $B^{-}\to \pi^{-}\mu^{+}\mu^{-}$ process and investigate the Majorana
neutrino contribution to its decay width. The constraints on the Majorana neutrino
mass and mixing parameter are obtained from this decay channel with the latest LHCb data.
Utilizing the best fit for the parameters, we study the lepton number violating decay
$B^{-}\to \pi^{+}\mu^{-}\mu^{-}$, and find its branching ratio is about $6.4\times10^{-10}$,
which is consistent with the LHCb data reported recently.
\end{abstract}
\end{frontmatter}
\section{Introduction}\label{intro}
In standard model (SM), neutrinos only take part in the weak interaction
and are massless particles. However, the neutrino oscillations discovered
in neutrino experiments indicate that neutrinos have non-zero mass
\cite{KamLAND,SNO,NEMO,Barger,Nmass1301.1340}.
The $10^{12}$ order hierarchy in $m_{t}/m_{\nu}$ and the large mixing
in neutrino sector suggest a possible new mechanism for neutrino mass generation
which is different from the fermion mass generated in SM such as
seesaw mechanisms \cite{see-saw}. In type I seesaw mechanism, the neutrino mass is generated by introducing heavy right-handed
Majorana neutrino. The Majorana nature of massive neutrinos typically manifests itself
in some processes where the lepton number can be violated by two units.
In order to disentangle the neutrino mass generation mechanism,
many efforts have been made to study such lepton number violating (LNV) processes
\cite{Si0810.5266,Han1211.6447,Castro1212.0037,WeiChao0804.1265,WeiChao0907.0935}.

It is well known that neutrinoless nuclear double beta ($0\nu\beta\beta$) decays
can be induced by Majorana neutrino. For the exchange of a light Majorana neutrino,
$0\nu\beta\beta$ decay rate is proportional to the square of the effective Majorana mass,
and many strong constraints have been obtained under this assumption
\cite{9808367,9902014,1305.3306}. For the heavy Majorana neutrino, the heavy meson
LNV decays have been studied extensively both in theory and at experiment.
Many LNV processes from pseudoscalar meson and $\tau$ lepton decays are studied,
and the existing limits on the mass and mixing of the heavy Majorana neutrino
are extracted from the experimental data in \cite{Han0901.3589}.
The rare decays of heavy mesons to a vector or pseudoscalar meson
are investigated in \cite{G.L.Wang1,S.S.Bao1208.5136}.
The four-body LNV decays of heavy pseudoscalar B and D mesons are studied
in \cite{G.L.Wang2,1108.6009}.
The upper limits for the branching fractions of the B meson LNV decays have been obtained.
$B^{+}\to D^{-}e^+e^+(e^+\mu^+,\mu^+\mu^+)$ are measured by Belle collaboration \cite{Belle},
$B^{+}\to \pi^{-}(K^{-})e^+e^+(\mu^+\mu^+)$ by BaBar collaboration \cite{BABAR},
and $B^{-}\to D^{+}(D^{*+},\pi^{+},D_{s}^{+},D^{0}\pi^{+})\mu^-\mu^-$
by LHCb collaboration \cite{LHCb,LHCbMajoranaN}.
According to LHCb results, the upper limit for the branch ratio of $B^{-}\to \pi^{+}\mu^{-}\mu^{-}$
is $4.0\times 10^{-9}$ at $95\%$ confidence level (C.L.) which is applicable for
the heavy Majorana neutrino lifetime $\tau_{N}\lesssim 1~\mathrm{ps}$,
and a model dependent upper limit on the coupling between muon and a possible
Majorana neutrino as a function of $m_{N}$ ($250\mathrm{MeV}<m_{N}<5000\mathrm{MeV}$)
is also given for $\tau_{N}$ up to $1000~\mathrm{ps}$ at $95\%$ C.L..

Recently, $B^{-}\to\pi^{-}\mu^{+}\mu^{-}$ has been observed with high statistics
at the LHCb experiment, with its branching ratio measured at
$(2.3\pm0.6(stat.)\pm0.1(syst.))\times10^{-8}$ ($5.2\sigma$ significance) \cite{Bpill-LHCb}.
This rare decay is induced by flavor changing neutral current $b\to d\ell^{+}\ell^{-}$
and suppressed in SM. This kind of process can be used to probe the new physics
beyond SM. Many theoretical studies on this channel within SM have been porformed
\cite{BPill-Wang0711.0321,BPill-Xiao1207.0265,BPill-Ali1312.2523,Bpill-Faustov1403.4466},
and new physics contributions can not be ruled out.
In this paper, we investigate the Majorana neutrino contribution to
$B^{-}\to \pi^{-}\mu^{+}\mu^{-}$, and obtain constraints on the mixing parameter
between muon and the Majorana neutrino.
Furthermore, we study the LNV decay $B^{-}\to \pi^{+}\mu^{-}\mu^{-}$
with the obtained mixing parameter, and our prediction for the
$B^{-}\to \pi^{+}\mu^{-}\mu^{-}$ branching ratio agrees with LHCb data.

This paper is organized as follows. In Sec.\ref{theo},
we briefly review the theoretical framework related to Majorana neutrino.
In Sec.\ref{num}, the numerical results and discussion are given.
Finally, we give a short summary.

\section{Theoretical Framework}\label{theo}

In one of the simplest extensions of the SM, Majorana neutrino mass term
is generated by introducing $n$ right-handed $\mathrm{SU(2)_{L}\times U(1)_{Y}}$
singlet neutrinos $N_{R}$ in addition to the three generations of left-handed
$\mathrm{SU(2)_{L}}$ doublet leptons $L_{L}$ and right-handed charged leptons $\ell_{R}$,
\begin{equation}
L_{L} =\left(\begin{array}{c} \nu_{\ell} \\ \ell
\end{array}\right)_{L},~~~\ell_{R},~~~N_{R}.
\end{equation}
The Dirac mass terms are generated with the Yukawa couplings to the Higgs doublet
in the SM, and the corresponding gauge invariance allows the right-handed neutrino singlets
to have Majorana mass terms.
The gauge-invariant Lagrangian relevant to lepton masses is expressed as
\begin{eqnarray}
-\mathcal{L}_{Y}=Y_{\ell}\overline{L_{L}}H\ell_{R}+Y_{\nu}\overline{L_{L}}\tilde{H}N_{R}
+\overline{N_{R}^{c}}M_{R}N_{R}+h.c.
\end{eqnarray}
where $H$ denotes the Higgs doublet, and $\tilde{H}=i\sigma_{2}H^{*}$.
$M_{R}$ is the right-handed Majorana neutrino mass matrix.
After the spontaneous gauge symmetry breaking, the mass matrix of charged leptons
$M_{\ell}=Y_{\ell}\,v/\sqrt{2}$ and the Dirac neutrino mass matrix $M_{D}=Y_{\nu}v/\sqrt{2}$
are obtained, with the vacuum expectation value $\langle H\rangle=v/\sqrt{2}$.
The complete neutrino mass sector is composed of both Dirac and Majorana mass term
\begin{eqnarray}
-\mathcal{L}_{M}=\left(\begin{array}{c}\overline{\nu_{L}},\overline{N_{R}^{c}}\end{array}\right)
\left(\begin{array}{c}~~0~~~M_{D} \\ M_{D}^{T}~~M_{R}\end{array}\right)
\left(\begin{array}{c} \nu_{L}^{c} \\ N_{R} \end{array}\right)+h.c.
\end{eqnarray}
At the leading-order in $M_{R}^{-1}$, mass matrix
for the three light neutrinos can be written as
\begin{eqnarray}
M_{\nu}\thicksim-M_{D}M_{R}^{-1}M_{D}^{T},
\end{eqnarray}
This is the Type I seesaw mechanism
which connects the small neutrino masses to the heavy Majorana neutrino masses.
Seesaw mechanism provides a natural explanation for the tiny neutrino mass.
The gauge interaction Lagrangian for the charged current in terms of
the neutrino mass eigenstates has the following form
\begin{eqnarray}
-\mathcal{L}=\frac{g}{\sqrt{2}}W_{\mu}^{+}
\Big(\sum_{\ell=e}^{\tau}\sum_{m=1}^{3}U_{\ell m}^{*}\overline{\nu_{m}}\gamma^{\mu}P_{L}\ell
+\sum_{\ell=e}^{\tau}\sum_{m^{'}=4}^{3+n}V_{\ell m^{\prime}}^{*}\overline{N_{m^{\prime}}^{c}}\gamma^{\mu}P_{L}\ell\Big)+h.c.
\end{eqnarray}
where $P_{L}=(1-\gamma_{5})/2$, $\nu_{m}(m=1,2,3)$ and $N_{m^{\prime}}(m^{\prime}=4,\cdots,3+n)$
are the mass eigenstates, $U_{\ell m}$ ($V_{\ell m^{\prime}}$)
is the mixing matrix element between the lepton flavor and light (heavy) neutrinos.
In this papar, we simply assume that there is only one heavy Majorana neutrino $N$,
with its mass $m_{\pi}<m_{N}<m_{B}$ where $m_{B(\pi)}$ is the mass of the B ($\pi$) meson.


\begin{figure}[!htb]
\begin{center}
\subfigure[]{ \label{fig:Ma}
\includegraphics[scale=0.4]{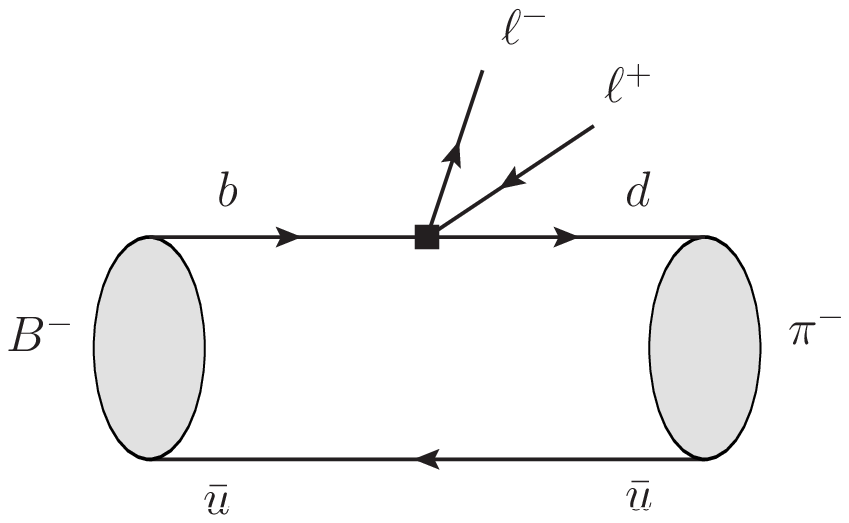}
}
\subfigure[]{ \label{fig:Mb}
\includegraphics[scale=0.4]{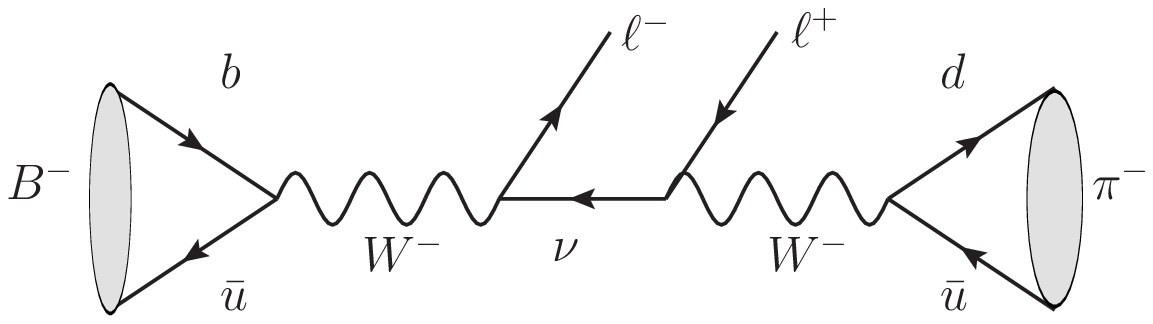}
}
\subfigure[]{ \label{fig:Mc}
\includegraphics[scale=0.4]{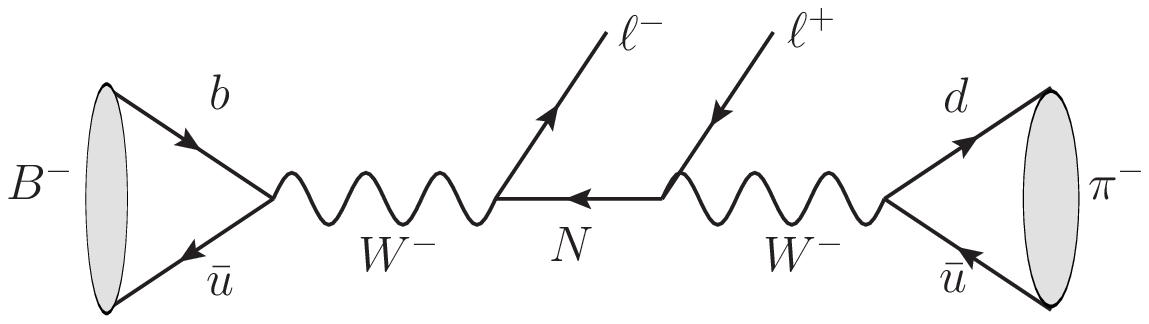}
}
\caption{Feynman diagrams for the decay process $B^{-}\to \pi^{-}\ell^{+}\ell^{-}$.}\label{fig:2}
\end{center}
\end{figure}

First, we study the process $B^{-}(p)\to \ell^{-}(p_{1})\ell^{+}(p_{2})\pi^{-}(p_{3})$,
where $p$, $p_{1}$, $p_{2}$ and $p_{3}$ denote the four-momentum of the corresponding particle.
The dominated Feynman diagram in SM is displayed in Fig.\ref{fig:Ma},
where the black square denotes the effective vertex of
the leading-order $b\to d\ell^{+}\ell^{-}$ transition \cite{LuCaiDian}.
The corresponding amplitude can be written as
\begin{align}
\mathcal{M}_{a}=&\frac{G_{F}\alpha}{\sqrt{2}\pi}V_{tb}V_{td}^{*}[C_{9}^{eff}\langle\pi(p_{3})|\bar{d}\gamma_{\mu}P_{L}b|B(p)\rangle\bar{u}(p_{1})\gamma^{\mu}v(p_{2})\nonumber\\
&+C_{10}^{eff}\langle \pi(p_{3})|\bar{d}\gamma_{\mu}P_{L}b|B(p)\rangle\bar{u}(p_{1})\gamma^{\mu}\gamma_{5}v(p_{2})\nonumber\\
&-2C_{7}^{eff}\frac{1}{q^{2}}\langle \pi(p_{3})|\bar{d}i\sigma_{\mu\nu}q^{\nu}(m_{b}P_{R}+m_{d}P_{L})b|B(p)\rangle\bar{u}(p_{1})\gamma^{\mu}v(p_{2})],\label{eq:amplitude1}
\end{align}
with $P_{L,R}=(1\mp\gamma_{5})/2$ and $s=q^{2}=(p_{1}+p_{2})^{2}=(p-p_{3})^{2}$.
$m_{b(d)}$ denotes the mass of the $b$ ($d$) quark.
$G_{F}$ is the Fermi coupling constant. $\alpha$ is the fine-structure constant
and $V_{q_{1}q_{2}}$ is the CKM matrix element.
The analytic expressions for all Wilson coefficients $C_{i}$, except $C_{9}^{eff}$,
are the same as that used to study the $b\to s$ transition, and can be founded in \cite{Buras95Ci}.
The next-to-leading approximation for Wilson coefficient $C_{9}^{eff}$
can be written as \cite{CP-BXsll}
\begin{align}
C_{9}^{eff}=&C_{9}[1+ \frac{\alpha_{s}(\mu)}{\pi}\omega(\hat{s})]
+g(\hat{m}_{c},\hat{s})(3C_{1}+C_{2}+3C_{3}+C_{4}+3C_{5}+C_{6})\nonumber\\
&-\frac{1}{2}g(\hat{m}_{d},\hat{s})(C_{3}+3C_{4})
-\frac{1}{2}g(\hat{m}_{b},\hat{s})(4C_{3}+4C_{4}+3C_{5}+C_{6})\nonumber\\
&+\frac{2}{9}(3C_{3}+C_{4}+3C_{5}+C_{6})
+\lambda_{u}[g(\hat{m}_{c},\hat{s})-g(\hat{m}_{u},\hat{s})](3C_{1}+C_{2}),
\end{align}
where $\lambda_{u}=V_{ub}V_{ud}^{*}/V_{tb}V_{td}^{*}\,$,
$\hat{m}_{j}=m_{j}/m_{B}$ with $j=u,d,c,b$ and $\hat{s}=s/m_{B}^{2}$.
The values of Wilson coefficients $C_{i}$ ($i=1,2,\cdots,10$) used in this work
are listed in Tab.\ref{tab:WilsonC}.
The relevant expressions of $\omega(\hat{s})$ and $g(\hat{m}_{j},\hat{s})$ can be found in \cite{CP-BXsll,CP-BPi}.
The hadronic matrix elements in the decay amplitude can be parameterized
in terms of $B\to\pi$ form factors $f_{+}^{B\pi}(q^{2})$,
$f_{0}^{B\pi}(q^{2})$ and $f_{T}^{B\pi}(q^{2})$,
\begin{eqnarray}
&&\langle \pi(p_{3})|\bar{d}\gamma_{\mu}b|B(p)\rangle
=(p+p_{3})_{\mu}f_{+}^{B\pi}(q^{2})
+\frac{m_{B}^{2}-m_{\pi}^{2}}{q^{2}}q_{\mu}\big(f_{0}^{B\pi}(q^{2})-f_{+}^{B\pi}(q^{2})\big),\nonumber\\
&&\langle \pi(p_{3})|\bar{d}\sigma_{\mu\nu}q^{\nu}b|B(p)\rangle
=i[(p+p_{3})_{\mu}q^{2}-(m_{B}^{2}-m_{\pi}^{2})q_{\mu}]\frac{f_{T}^{B\pi}(q^{2})}{m_{B}+m_{\pi}}.\label{eq:FormFactors}
\end{eqnarray}

\begin{table}[h]
\centering \caption {The values of Wilson coefficients $C_{i}(\mu)$ in SM at the scale $\mu=m_{b}$
at the leading logarithmic approximation, with $m_{W}=80.4~\mathrm{GeV}$,
$m_{t}=173.5~\mathrm{GeV}$, $m_{b}=4.8~\mathrm{GeV}$.}\label{tab:WilsonC}
\begin{tabular}{cccccccccc}
\hline
 $C_{1}$ &$C_{2}$ &$C_{3}$ &$C_{4}$ &$C_{5}$
&$C_{6}$ &$C_{7}$ &$C_{8}$ &$C_{9}$ &$C_{10}$\\
\hline
-0.246 &1.106 &0.011 &-0.025 &0.007
&-0.031 &-0.312 &-0.187 &4.211 &-4.501 \\
\hline
\end{tabular}
\end{table}

Fig.\ref{fig:Mc} can have an enhanced contribution when the intermediate Majorana neutrino is onshell.
The amplitudes corresponding to Fig.\ref{fig:Mb} and \ref{fig:Mc} are as follows
\begin{eqnarray}
\mathcal{M}_{b}&=&8G_{F}^{2}V_{ub}V_{ud}^{*}\frac{\bar{u}(p_{1})\gamma_{\mu}p\!\!\!/_{\nu}\gamma^{\rho}P_{L}v(p_{2})}{p_{\nu}^{2}}\nonumber\\
&&\times\langle\pi(p_{3})|\bar{d}\gamma_{\rho}P_{L}u|0\rangle\langle 0|\bar{u}\gamma^{\mu}P_{L}b|B(p)\rangle,\label{eq:amplitude2}\\
\mathcal{M}_{c}&=&8G_{F}^{2}V_{ub}V_{ud}^{*}V_{\ell_{1}N}V_{\ell_{2}N}
\frac{\bar{u}(p_{1})\gamma_{\mu}p\!\!\!/_{N}\gamma^{\rho}P_{L}v(p_{2})}{p_{N}^{2}-m_{N}^{2}+im_{N}\Gamma_{N}}\nonumber\\
&& \times \langle\pi(p_{3})|\bar{d}\gamma_{\rho}P_{L}u|0\rangle\langle 0|\bar{u}\gamma^{\mu}P_{L}b|B(p)\rangle,\label{eq:amplitude3}
\end{eqnarray}
where $p_{\nu(N)}$ represents the four-momentum of
the light (heavy) neutrino $\nu$ ($N$).
$V_{\ell N}$ is the mixing matrix element between the lepton flavor $\ell$
and heavy neutrino $N$. The hadronic matrix elements in eq.\eqref{eq:amplitude2} and \eqref{eq:amplitude3}
are expressed as
\begin{eqnarray}
&&\langle 0|\bar{u}\gamma^{\mu}(1-\gamma_{5})b|B(p)\rangle=-if_{B}p^{\mu},\nonumber\\
&&\langle\pi(p_{3})|\bar{d}\gamma^{\mu}(1-\gamma_{5})u|0\rangle=if_{\pi}p_{3}^{\mu}.
\end{eqnarray}
where $f_{B(\pi)}$ is the decay constant of the B ($\pi$) meson.
$\Gamma_{N}$ is the total decay width of
the Majorana neutrino, and can be estimated by \cite{G.Cvetic}
\begin{eqnarray}
\Gamma_{N}\approx 2\sum_{\ell}|V_{\ell N}|^{2}\Big(\frac{m_{N}}{m_{\tau}}\Big)^{5}\times \Gamma_{\tau},
\end{eqnarray}
where $m_{\tau}$ and $\Gamma_{\tau}$ denote the mass and total decay width of
the $\tau$ lepton, respectively.
The decay branching ratio of $B^{-}\to \pi^{-}\ell^{+}\ell^{-}$ is given by
\begin{eqnarray}
\mathcal{B}
=\frac{\tau_{B}}{2m_{B}(2\pi)^{5}}\int|\mathcal{M}_{a}+\mathcal{M}_{b}+\mathcal{M}_{c}|^{2}\frac{|\vec{p}_{1}^{B}|}{4m_{B}}
\frac{|\vec{p}_{2}^{*}|}{4\sqrt{s_{23}}}~d\Omega_{1}^{B}d\Omega_{2}^{*}ds_{23},\label{eq:BranchingRatio}
\end{eqnarray}
where $\vec{p}_{1}^{\,B}$ ($\vec{p}_{2}^{\,\,*}$) and $d\Omega_{1}^{B}$ ($d\Omega_{2}^{*}$)
denote the 3-momentum and solid angle of charged lepton $\ell^{-}$ ($\ell^{+}$)
in the rest frame of B meson ($\ell^{+}\pi$ system), respectively.
It is found that the contributions from Fig.\ref{fig:Mb}
and from the interferences can be neglected.


\begin{figure}[!ht]
\begin{center}
\subfigure[]{ \label{fig:Na}
\includegraphics[scale=0.5]{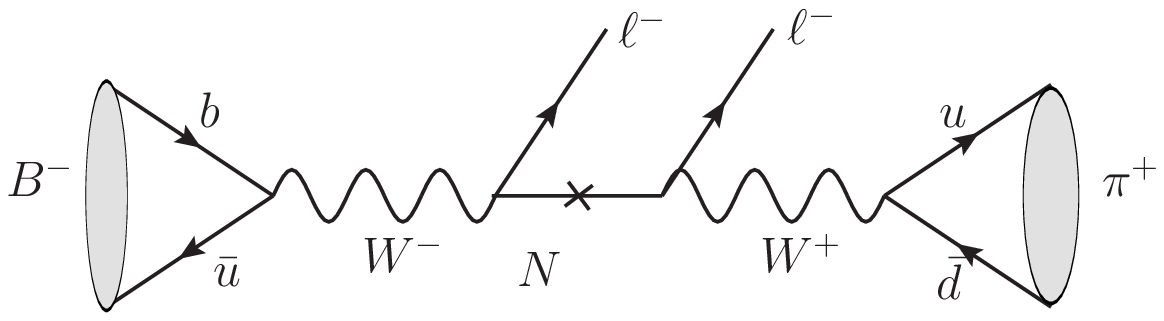}
}
\subfigure[]{ \label{fig:Nb}
\includegraphics[scale=0.5]{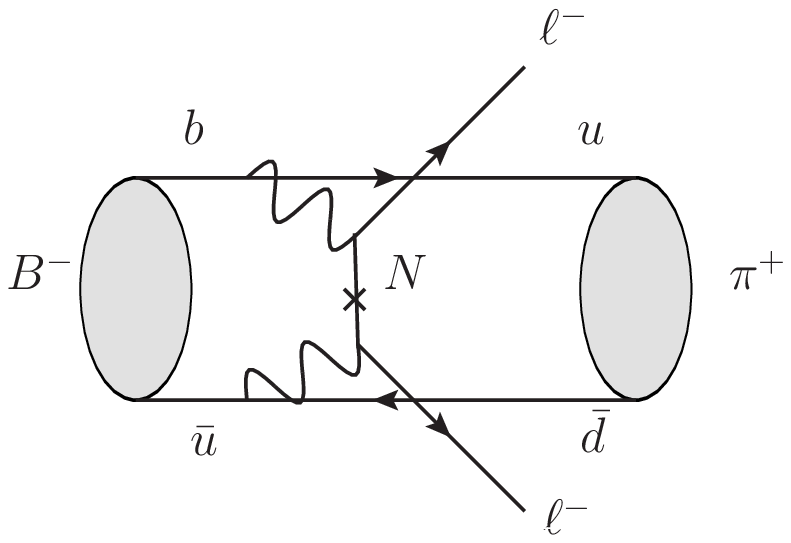}
}
\caption{Feynman diagrams for $B^{-}\to \pi^{+}\ell^{-}\ell^{-}$ via
Majorana neutrino exchange.}\label{fig:N}
\end{center}
\end{figure}

Next, we study the LNV process $B^{-}(p)\to \ell^{-}(p_{1})\ell^{-}(p_{2})\pi^{+}(p_{3})$.
Such $\Delta L=2$ process may occur via Majorana neutrino exchange.
In this case, this process is dominated by the annihilation diagram shown in Fig.\ref{fig:Na}
as the intermediate neutrino can be on-shell and has a resonantly enhanced effect,
while the contribution from the emission diagram in Fig.\ref{fig:Nb}
is small enough and can be neglected \cite{S.S.Bao1208.5136}.
Omitting the light charged lepton mass, one can obtain the corresponding
decay branching ratio as follows
\begin{eqnarray}
\mathcal{B}(B^{-}\to \pi^{+}\ell^{-}\ell^{-})
&=&\frac{\tau_{B}G_{F}^{4}f_{B}^{2}f_{\pi}^{2}}{128\pi^{2}}
|V_{ub}V_{ud}^{*}|^{2}|V_{\ell N}|^{4}
\frac{m_{B}m_{\tau}^{5}}{2\Gamma_{\tau}}\nonumber\\
&&\times
\left(1-\frac{m_{\pi}^{2}}{m_{N}^{2}}\right)^{2}
\left(1-\frac{m_{N}^{2}}{m_{B}^{2}}\right)^{2}.
\end{eqnarray}

\section{Numerical Analysis}\label{num}

\begin{table}[ht]
\centering \caption {The form factors obtained by light cone sum rule and the fitted parameters for
z-series parameterization \cite{preparation}.}\label{tab:FormFactors}
\begin{tabular}{cccccc}
\hline
 $f_{+}^{B\pi}(0)$ &$b_{1}^{+}$~~~~
 &$f_{0}^{B\pi}(0)$ &$b_{1}^{0}$~~~~
 &$f_{T}^{B\pi}(0)$ &$b_{1}^{T}$\\
\hline
0.275 &-2.037~~~~ &0.275 &-2.808~~~~ &0.293 &-1.780 \\
\hline
\end{tabular}
\end{table}

In order to estimate the SM contribution as precisely as possible,
we adopt the following simplified Boyd-Grinstein-Lebed (BGL) version \cite{BGL9412324}
of z-series parameterization forms,
\begin{eqnarray}
f_{+(T)}^{B\pi}(q^{2})&=&\frac{f_{+(T)}^{B\pi}(0)}{1-q^{2}/m_{B^{*}}^{2}}
\bigg\{1+b_{1}^{+(T)}\Big[z(q^{2},t_{0})-z(0,t_{0})\nonumber\\\label{eq:f+T}
&&+\frac{1}{2}\Big(z(q^{2},t_{0})^{2}-z(0,t_{0})^{2}\Big)\Big]\bigg\},\\
f_{0}^{B\pi}(q^{2})&=&f_{0}^{B\pi}(0)
\bigg\{1+b_{1}^{0}\Big(z(q^{2},t_{0})-z(0,t_{0})\Big)\bigg\},\label{eq:f0}
\end{eqnarray}
where $m_{B^{*}}=5.325~\mathrm{GeV}$ denotes the mass of the vector meson $B^{*}$.
$f_{+(T)}^{B\pi}(0)$ and $f_{0}^{B\pi}(0)$ are $B\to\pi$ form factors at $q^{2}=0$.
These form factors can be obtained by the light cone sum rules.
The corresponding values are listed in Tab.\ref{tab:FormFactors}.
The function $z(q^{2},t_{0})$ has the following form,
\begin{eqnarray}\label{eq:Zparameter}
z(q^{2},t_{0})=\frac{\sqrt{t_{+}-q^{2}}-\sqrt{t_{+}-t_{0}}}{\sqrt{t_{+}-q^{2}}+\sqrt{t_{+}-t_{0}}},
\end{eqnarray}
with the auxiliary parameters $t_{\pm}=(m_{B}\pm m_{\pi})^{2}$ and
$t_{0}=t_{+}-\sqrt{(t_{+}-t_{-})(t_{+}-q_{min}^{2})}$.
In our numerical calculations,
the CKM matrix elements are obtained by the Wolfenstein parametrization with
$\lambda=0.22535$, $A=0.817$, $\bar{\rho}=0.136$ and $\bar{\eta}=0.348$.
The other parameters used in this paper are collected in Tab.\ref{tab:BpillInPut}.

\begin{table}[ht]
\begin{center}
\caption{Parameters used in our numerical calculation and the values taken from PDG \cite{PDG2012}.}
\label{tab:BpillInPut}
\begin{tabular}{cccc}
\hline
    $\alpha$               & $1/137$                                     &$m_{u}$         & $2.3~\mathrm{MeV}$\\
    $G_{F}$                & $1.16637\times10^{-5}~\mathrm{GeV}^{-2}$    &$m_{d}$         & $4.8~\mathrm{MeV}$\\
    $\tau_{B}$             & $1.641~\mathrm{ps}$                         &$m_{c}$         & $1.275~\mathrm{GeV}$\\
    $\Gamma_{\tau}$        & $2.3\times10^{-12}~\mathrm{GeV}$            &$m_{\tau}$      & $1.777~\mathrm{GeV}$\\
    $f_{\pi}$              & $130.4~\mathrm{MeV}$                        &$m_{\pi}$       & $0.13957~\mathrm{GeV}$\\
    $f_{B}$                & $194~\mathrm{MeV}$                          &$m_{B}$         & $5.279~\mathrm{GeV}$\\
\hline
\end{tabular}
\end{center}
\end{table}

\begin{figure}[!ht]
\begin{center}
\includegraphics[scale=0.8]{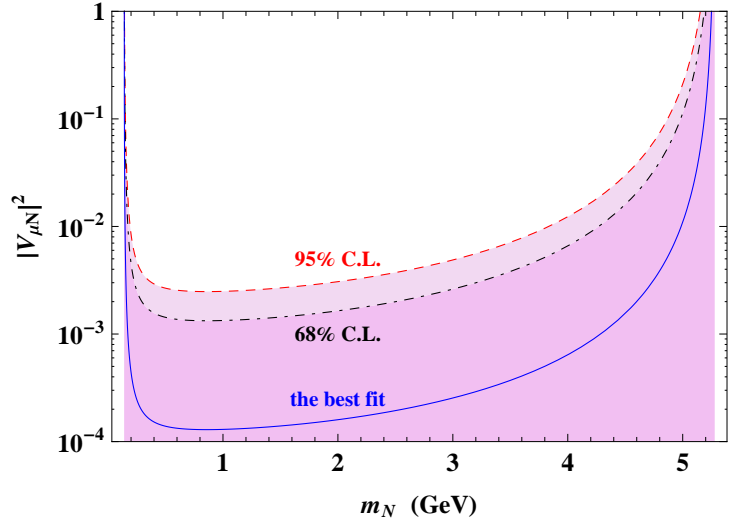}
\caption{Contour plot obtained from
$B^{-}\to \pi^{-}\mu^{+}\mu^{-}$ for the Majorana Neutrino mass and the mixing parameter between the light flavour $\mu$
and the Majorana neutrino $N$. }\label{fig:Vln}
\end{center}
\end{figure}

By analyzing $B^{-}\to \pi^{-}\mu^{+}\mu^{-}$ process and comparing
our numerical results with LHCb experiment data, we obtain the contour plot
for the Majorana neutrino mass $m_N$ and the $V_{\mu N}$ which is the mixing parameter between $\mu$ and the Majorana neutrino $N$.
The results are displayed in Fig.\ref{fig:Vln}.
The region above (below) the dashed line is excluded (allowed) by LHCb
with $B\to\pi\mu^{+}\mu^{-}$ at $95\%$ C.L..
The dot-dashed line is the boundary at $68\%$ C.L..
The solid line denotes the best fit for $|V_{\mu N}|^{2}$ and $m_{N}$.
One can notice from the solid line that at $m_{N}=3~\mathrm{GeV}$,
$|V_{\mu N}|^{2}\approx 2.5\times10^{-4}$.
Using the best fit for $|V_{\mu N}|^{2}$ and $m_{N}$,
we have the branching ratio for the LNV decay $B^{-}\to \pi^{+}\mu^{-}\mu^{-}$,
\begin{eqnarray}
\mathcal{B}(B^{-}\to \pi^{+}\mu^{-}\mu^{-})=6.5\times10^{-10}.
\end{eqnarray}
This result agrees with LHCb data \cite{LHCbMajoranaN}
\begin{eqnarray}
\mathcal{B}_{exp}(B^{-}\to \pi^{+}\mu^{-}\mu^{-})\leq 4.0\times10^{-9}
~~~~~\text{at $95\%$ C.L.}
\end{eqnarray}
The CP violation effect is too small to be observed, so that
$\mathcal{B}(B^{+}\to \pi^{-}\mu^{+}\mu^{+})=
\mathcal{B}(B^{-}\to \pi^{+}\mu^{-}\mu^{-})=6.5\times10^{-10}$.

\section{Summary}\label{sum}
The Type I seesaw mechanism is one of
the natural schemes to describe the tiny neutrino mass. In this paper,
we study the contribution from Majorana neutrino to the semi-leptonic B-meson rare decays
within Type I seesaw mechanism.
The constraints for the Majorana neutrino parameters are obtained by analyzing
$B^{-}\to \pi^{-}\mu^{+}\mu^{-}$ process from the latest LHCb data.
It is found that new physics effects from Majorana neutrino cannot be eliminated.
Then we adopt these parameters to investigate the LNV B-meson decay
$B^{-}\to \pi^{+}\mu^{-}\mu^{-}$. Our result for its branching ratio
is consistent with the upper limit given by LHCb experiment.
The B meson rare decays can be measured with high precision at LHCb in the near future.
As a result, it is possible to search for new physics by studying the B meson rare decays.
In particular, studying the LNV B-meson decays will be important
to explore the neutrino mass mechanism.


\section*{Acknowledgments}
The authors would like to thank Dr. X. Gong and Prof. S. Y. Li for the helpful discussions.
This work was supported in part by NSFC under grant Nos.11325525, 112755114 and 11235005,
and NSF of Shandong Province.

\section*{References}
\bibliographystyle{elsarticle-num}

\end {document}